\documentclass[pre,aps,floatfix,superscriptaddress,two column]{revtex4-2}
\usepackage{graphicx}
\usepackage{dcolumn}
\usepackage{latexsym}
\usepackage{hyperref}
\usepackage{amsmath, amsthm, amssymb}
\usepackage{relsize}
\usepackage{epsfig}
\usepackage{bm}
\usepackage{geometry}
\usepackage{xcolor}
\usepackage{placeins}
\usepackage[normalem]{ulem}
\usepackage{diagbox}
\usepackage[normalem]{ulem}
\usepackage{tikz}
\usepackage{multirow}
\usepackage{makecell}

\renewcommand{\arraystretch}{1.5}

\usepackage{titlesec}

\titlespacing*{\section}
  {0pt}{*2}{*0.5}  

\titlespacing*{\subsection}
  {0pt}{*1.5}{*0.3}

\begin{document}


\title{Data-Driven Equation Discovery for Phase-Ordering Dynamics: From Allen–Cahn to the Ising Model}

\author{Partha Sarathi Mondal}
\email[]{parthasarathimondal.rs.phy21@itbhu.ac.in}
\affiliation{Department of Physics, Indian Institute of Technology (BHU) Varanasi, India 221005}
\author{Manav Kumar Jalan}
\email[]{manav.krjalan.cd.phy23@itbhu.ac.in}
\affiliation{Department of Physics, Indian Institute of Technology (BHU) Varanasi, India 221005}
\author{Anish Kumar}
\email[]{anishkumar.rs.phy22@itbhu.ac.in}
\affiliation{Department of Physics, Indian Institute of Technology (BHU) Varanasi, India 221005}
\author{Shradha Mishra}
\email[]{smishra.phy@itbhu.ac.in}
\affiliation{Department of Physics, Indian Institute of Technology (BHU) Varanasi, India 221005}




\begin{abstract}
Data-driven discovery of governing equations from spatiotemporal data offers a promising route to obtaining coarse-grained descriptions of complex dynamical systems. Here, we investigate the performance of PDE-SINDy for discovering phase-ordering dynamics using the Allen--Cahn equation as a benchmark and the Ising model with Glauber spin-flip dynamics as a microscopic system. We systematically analyze the effects of data availability, size of the candidate library, and noise on the efficiency of the equation discovery. We find that stability-selection PDE-SINDy can robustly identify the relevant terms in the governing dynamics even under limited or noisy data, while the recovered coefficient values are substantially more sensitive to these factors. We further show that enlarging the candidate library can strongly affect both term identification and coefficient recovery. Incorporating library bagging with stability selection reduces this sensitivity and improves the efficiency of equation discovery. For the Glauber spin flip Ising model dynamics, the resulting coarse-grained equation reproduces the characteristic phase-separation and coarsening dynamics of the underlying microscopic system. Overall, our results demonstrate the potential of PDE-SINDy for phase-ordering systems while highlighting the importance of carefully assessing the factors that influence the efficiency of equation discovery.

\end{abstract}

\maketitle
\section{Introduction}\label{sec:intr}
In many-body systems, interactions among microscopic constituents give rise to collective phenomena, such as phase separation, pattern formation, and flocking, that emerge on length and time scales much larger than those of the individual constituents \cite{ChaikinLubensky1995,cross1993,hohenberg1977theory}. Mesoscopic field-theoretic descriptions provide a powerful framework for understanding the kinetics and physical mechanisms underlying such emergent behavior. By identifying a small number of relevant, slowly varying fields at mesoscopic scales \cite{forster2018hydrodynamic}, continuum descriptions retain the essential dynamical features of the underlying microscopic system while eliminating the microscopic degrees of freedom that are not directly relevant to the emergent behavior.\\
For complex systems, even though the relevant slow fields can often be identified, deriving the dynamical equations governing their evolution remains challenging. In a bottom-up approach \cite{dean1996langevin,bertin2006boltzmann}, one aims to derive the field theory by systematically coarse graining the underlying microscopic dynamics. Such derivations are often analytically intractable for complex systems and typically involve approximations that are system-specific, limiting their general applicability. Alternatively, a top-down approach \cite{toner1998flocks,ramaswamy2010mechanics,cates2015motility} constructs the governing equations phenomenologically based on symmetry principles and conservation laws. While broadly applicable, this approach also has important limitations: the coefficients are generally phenomenological, with their dependence on microscopic parameters often unclear, and more importantly, identifying the relevant terms requires prior physical insight into the system's dynamics.\\
Recent developments in machine learning (ML) have provided alternative approaches for inferring coarse-grained dynamics directly from data. These range from neural-network (NN) architectures that learn complex dynamical behaviors to symbolic regression methods that provide explicit mathematical representations of the governing equations \cite{brunton2016discovering,cranmer2023interpretable,udrescu2020ai}. While NN-based approaches can achieve high predictive accuracy, their black-box nature often limits the interpretability of the learned dynamics \cite{guidotti2018survey,molnar2022interpretable,rudin2019stop}. They also typically require large training datasets \cite{goodfellow2016deep,lecun2015deep,karniadakis2021physics} and may exhibit limited generalization beyond the training regime \cite{geirhos2020shortcut,recht2019imagenet,ovadia2019can,yang2021b}, which can be restrictive when data are limited. Symbolic regression, in contrast, provides explicit mathematical representations of the governing equations and can often operate effectively with comparatively smaller datasets. Among symbolic regression approaches, the Sparse Identification of Nonlinear Dynamics (SINDy) framework has emerged as a powerful method for discovering governing equations directly from data \cite{brunton2016discovering,rudy2017data,brunton2022data}. Assuming that the underlying dynamics can be represented as a sparse combination of terms drawn from a predefined library of candidate functions and formulates equation discovery as a sparse regression problem.\\
\begin{figure*}[hbt!]
    \centering
    \includegraphics[width=0.995\linewidth]{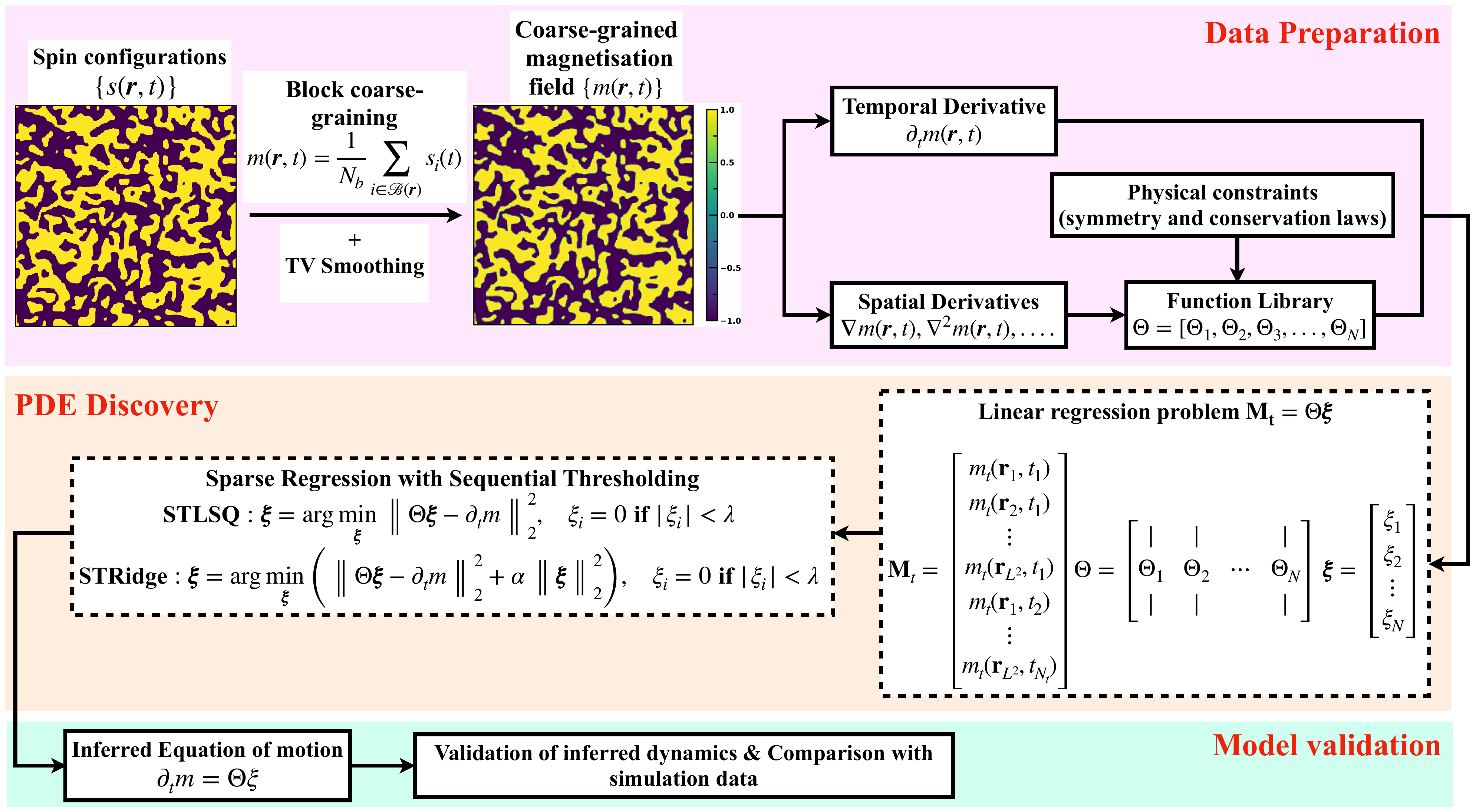}
    \caption{Schematic representation of the flow chart of the equation discovery process using PDE SINDy.}
    \label{fig:PDE_SINDy_Flowchart}
\end{figure*}
SINDy has been successfully applied to a broad range of dynamical systems, demonstrating its ability to recover governing equations directly from spatiotemporal data. These studies have also highlighted several factors that can influence the efficiency of the equation discovery process, including the amount and quality of available data, the size of the candidate library. However, equation discovery in phase-ordering systems presents additional challenges. In coarsening systems, the nontrivial dynamics is primarily localized near interfaces separating the coexisting phases, while the bulk of the domains remains comparatively inactive. As coarsening proceeds, the total interfacial area decreases with time, reducing the fraction of the system that directly samples the dynamics responsible for domain evolution.\\
In this work, we systematically assess the performance of the PDE-SINDy framework for equation discovery in phase-ordering systems. We first consider the Allen--Cahn equation, one of the simplest continuum descriptions of coarsening dynamics, as a controlled setting in which the governing equation is known a priori. The broad applicability of the Allen--Cahn model across diverse physical and biological phase-ordering systems makes it a useful benchmark for assessing the generalizability of data-driven equation-discovery methods for phase-ordering dynamics. Using this benchmark, we systematically examine how the accuracy of equation recovery is affected by the amount of available data, the size of the candidate library, and the presence of noise. To improve the robustness of term identification, we incorporate stability selection into the PDE-SINDy procedure and additionally use library bagging for an enlarged candidate library. We then turn to the Glauber spin-flip Ising model, where the dynamics are generated from microscopic Monte Carlo simulations and the coarse-grained equation is not known a priori.\\
The rest of the manuscript is organized as follows. In Sec.II, we describe the methodology adopted in this work. In Sec.III, we present and discuss our results. Finally, in Sec.IV, we summarize our findings and conclude with a brief discussion of possible future directions.

\begin{figure*}[hbt]
    \centering
    \includegraphics[width=0.999\linewidth]{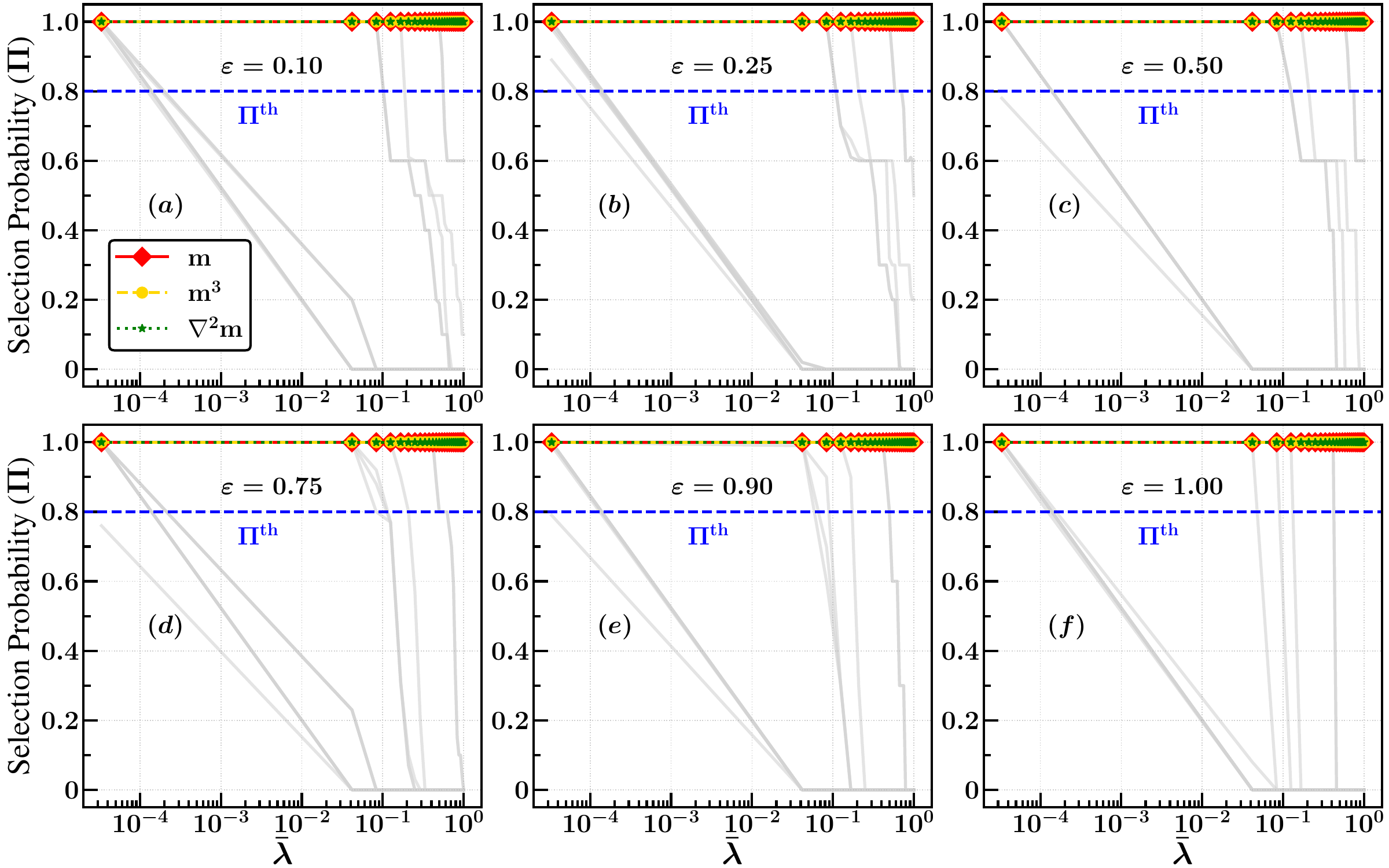}
    \caption{Selection probability ($\Pi$) of the candidate terms as a function of the normalized threshold parameter $\bar{\lambda}=\lambda/\lambda_{\max}$ for different fractions of the available data, $\varepsilon$. The three true terms in the Allen--Cahn equation, namely $m$, $m^3$, and $\nabla^2m$, are highlighted, while the remaining candidate terms are shown in gray. The horizontal dashed line denotes the selection probability threshold, $\Pi^{\mathrm{th}}=0.80$, used for model selection.}
    \label{fig:sel_prob_diff_data_frac_model}
\end{figure*}

\section{Methodology}\label{sec:meth}
We consider two prototypical systems, namely the Allen-Cahn equation and the two-dimensional Glauber spin flip Ising model. For both systems, the relevant slow variable is known \textit{a priori}, and is represented by a scalar field $m(\boldsymbol{r},t)$, corresponding to the scalar order parameter field and the local coarse-grained magnetization field, respectively. Our aim is to infer the corresponding governing continuum equations directly from spatiotemporal data using the PDE-SINDy framework \cite{rudy2017data}, an extension of the SINDy methodology to spatially extended dynamical systems. The equation-discovery procedure is summarized below.
\begin{itemize}
    \item{The first step is to obtain the coarse-grained magnetization field $m(\boldsymbol{r},t)$. For the Allen-Cahn (AC) system, the local magnetization field is directly obtained from the numerical solution of the governing PDE. For the Glauber spin flip Ising model (GIM), the field $m(\boldsymbol{r},t)$ is constructed by coarse-graining the spin configurations generated through Markov chain Monte Carlo simulations.}
    \item{The resulting spatiotemporal data for $m(\boldsymbol{r},t)$ are then used to construct the candidate function library and the corresponding temporal derivative vector.} 
    \item{Finally, the sparse regression is performed using the sequential threshold ridge regression (STRidge) algorithm to identify the relevant terms and their corresponding coefficients from the candidate function library. The governing equation is expressed in the form $\partial_t m=f(m)$.}
\end{itemize}
A schematic representation of the complete equation-discovery procedure is presented as a flowchart in Fig.\ref{fig:PDE_SINDy_Flowchart}. Details of the data generation for the AC and the GIM systems are provided in Appendix.\ref{app:data_gen_ac_gim}, while the details of the PDE-SINDy methodology are described in Appendix.\ref{app:pde_sindy}.

\section{Results}\label{sec:res}
\subsection{Allen-Cahn Equation}
We first systematically evaluate the performance of the PDE-SINDy framework by investigating the effects of data availability, candidate-library size, and noise on the recovery of the underlying governing equation for phase-ordering systems. As a benchmark, we consider the Allen--Cahn equation (Model A), which provides a canonical continuum description of phase-ordering dynamics governed by a single nonconserved order parameter, such as microstructure evolution in materials \cite{chen2002phase}, phase-ordering kinetics \cite{almeida2021phase}, solidification \cite{yaghi2022port}, fracture and crack propagation \cite{hu2020phase}, multicellular dynamics \cite{nonomura2012study}, image segmentation \cite{benevs2004geometrical}, and tumor growth \cite{gatti2025allen}. Its broad applicability in describing phase-ordering systems makes it a natural benchmark for assessing the ability of PDE-SINDy to recover nonlinear dynamical equations from spatiotemporal data for these systems.\\
The candidate library consists of polynomial terms in $m$ up to $m^6$, together with the spatial derivative terms $\nabla^2 m$ and $(\boldsymbol{\nabla}m)^2$. All possible products of the polynomial and derivative terms are considered, with the resulting terms restricted to at most second order in spatial gradients and sixth order in $m$. This procedure yields a library containing $18$ candidate functions, denoted by $\Theta_{18}$. Previous studies on data-driven equation discovery using SINDy have identified three key factors that can influence the accuracy of equation recovery: (i) the amount of available data, (ii) the size of the candidate library, and (iii) the presence of noise in the data. We therefore systematically investigate the influence of each of these factors on the recovery of the governing equation for the phase-ordering dynamics described by the Allen--Cahn equation. For cases (i) and (iii), the regression is performed using the function library $\Theta_{18}$.\\
For all cases, the data are obtained by numerically solving the Allen--Cahn equation on a $256\times256$ lattice. The complete dataset comprises $10$ independent simulation runs, with $500$ snapshots recorded at equally spaced time intervals from each run. The regression problem is then formulated from these spatiotemporal snapshots following the procedure described in Appendix.\ref{app:pde_sindy}.
\begin{figure*}[hbt]
    \centering
    \includegraphics[width=0.999\linewidth]{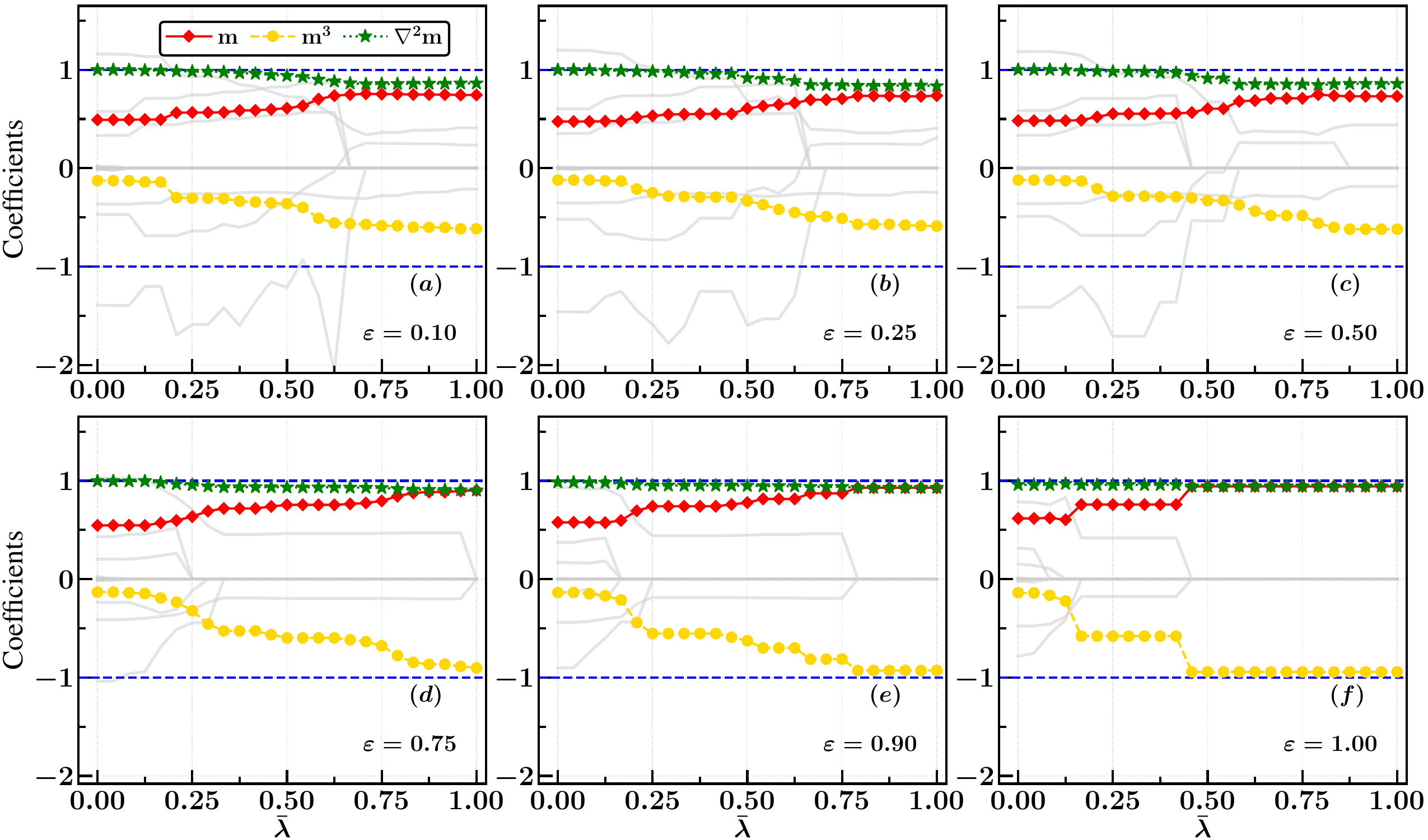}
    \caption{Recovered coefficients of the candidate terms as a function of the normalized threshold parameter $\bar{\lambda}=\lambda/\lambda_{\max}$ for different fractions of the available data, $\varepsilon$. The coefficients corresponding to the three true terms in the Allen--Cahn equation, namely $m$, $m^3$, and $\nabla^2m$, are highlighted, while the remaining candidate terms are shown in gray.}
    \label{fig:coeff_diff_data_frac_model}
\end{figure*}
\subsubsection{Effect of data size}
\begin{figure}
    \centering
    \includegraphics[width=0.999\linewidth]{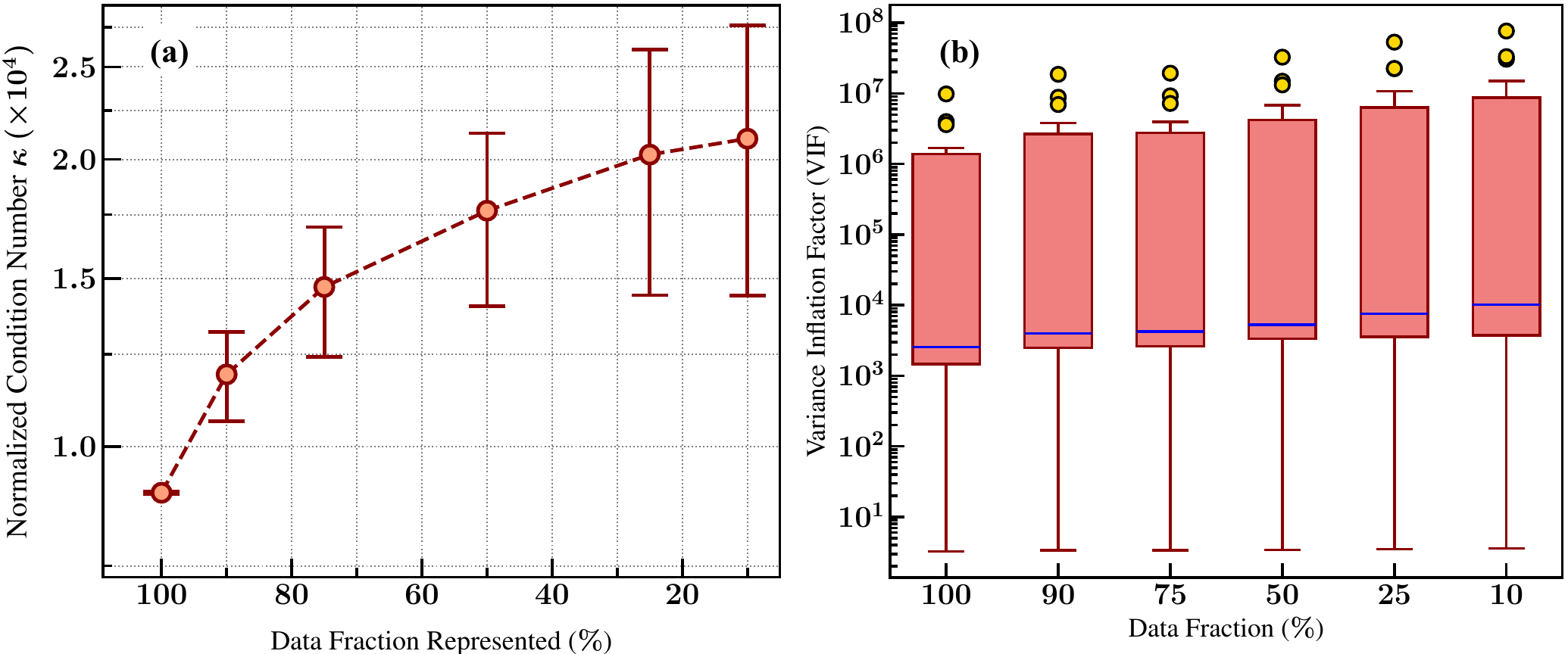}
    \caption{Variation of (a) the condition number, $\kappa$, of the library matrix and (b) the Variance Inflation Factor (VIF) of the candidate functions as a function of the available data fraction, $\varepsilon$. The box plots in (b) represent the distribution of VIF values for all candidate functions, with the horizontal line indicating the median, the box denoting the interquartile range, the whiskers showing the non-outlier range, and the circles representing outliers.}
    \label{fig:cond_VIF_librery}
\end{figure}

To investigate the effect of data availability on the accuracy of PDE-SINDy, we systematically vary the amount of spatiotemporal data used in the regression while keeping the underlying simulation parameters fixed. We characterize the data size by the fraction $\varepsilon$ of the complete dataset used for regression. The details of the data-sampling procedure are provided in Appendix.\ref{app:alg_det_data_sam}. We consider $\varepsilon \in \{0.10,0.25,0.50,0.75,0.90,1.0\}$ and compare the corresponding inferred equations to assess the robustness of equation recovery with respect to data availability.\\
We first examine the results of the stability-selection analysis. Fig.\ref{fig:sel_prob_diff_data_frac_model} shows the selection probability $\Pi$ of the different terms in the candidate library for different values of $\varepsilon$, plotted as a function of the normalized threshold parameter $\bar{\lambda}=\lambda/\lambda_{\max}\in[0,1]$. The three terms constituting the Allen--Cahn equation, namely $m$, $m^3$, and $\nabla^2m$, are retained with unit selection probability ($\Pi=1$) over the entire range of $\bar{\lambda}$ for all values of $\varepsilon$. Based on these results, we adopt a selection-probability threshold of $\Pi^{\mathrm{th}}=0.80$ to identify robust candidate terms for the inferred governing equation. For the complete dataset ($\varepsilon=1.0$), only these three terms satisfy this criterion for $\bar{\lambda}\geq0.50$, leading to the correct recovery of the governing equation. As the data fraction, $\varepsilon$, is reduced, several spurious terms acquire nonzero selection probabilities over a broader range of $\bar{\lambda}$. However, their selection probabilities remain below $\Pi^{th}$ and they are consequently excluded from the final model. These results demonstrate that reducing the amount of available data increases the susceptibility to spurious terms, but the correct governing equation can nevertheless be recovered through an appropriate stability-selection threshold.\\
Although the selection probabilities of the relevant terms remain largely insensitive to the amount of available data, the recovered coefficients exhibit a pronounced dependence on $\varepsilon$, as shown in Fig.~\ref{fig:coeff_diff_data_frac_model}. The results reveal three key features: (i) for all values of $\varepsilon$, the signs of the coefficients associated with the three terms of the Allen--Cahn equation are correctly recovered; (ii) the magnitudes of the recovered coefficients, however, depend strongly on $\varepsilon$. A mean-field analysis of Eq.~\ref{eq:allencahn} gives the equilibrium values of the order parameter inside the domains as $m_0=\pm\sqrt{-\alpha/\beta}$. For the data generation, we set $\alpha=1$ and $\beta=-1$, yielding $m_0=\pm1$. The results in Fig.~\ref{fig:coeff_diff_data_frac_model} show that the recovered coefficients reproduce the correct equilibrium values only for sufficiently large data fractions and sufficiently large values of $\bar{\lambda}$; (iii) the coefficient $\gamma$ of the $\nabla^2m$ term in Eq.~\ref{eq:allencahn} controls important interfacial properties, including the surface tension, interface width, and interface velocity during the phase separation process in Allen-Cahn equation. The recovered value of $\gamma$ deviates from its exact value of unity as $\varepsilon$ decreases. Nevertheless, the deviation remains relatively small, indicating that the resulting differences in the interfacial features is expected to be limited.\\
The above results demonstrate the distinct roles of data availability in the qualitative and quantitative aspects of equation discovery. While the correct terms are consistently identified across the entire range of $\varepsilon$, the quantitative accuracy of the recovered coefficients improves systematically with increasing data size.\\
Next, we examine the factors responsible for the reduced accuracy of the recovered dynamical equation as the data size decreases. To assess the conditioning of the resulting regression problem, we calculate the condition number of the library matrix,
\begin{equation}
\kappa(\Theta)=\frac{\sigma_{\max}}{\sigma_{\min}},
\end{equation}
where $\sigma_{\max}$ and $\sigma_{\min}$ denote the largest and smallest singular values of the library matrix $\Theta$, respectively. In addition, we calculate the Variance Inflation Factor (VIF) for each candidate function, defined as
\begin{equation}
\mathrm{VIF}_i=\frac{1}{1-R_i^2},
\end{equation}
where $R_i^2$ is the coefficient of determination obtained by regressing the $i^{\mathrm{th}}$ library term against all the remaining terms. While the condition number measures the overall conditioning of the regression problem, the VIF quantifies the degree of multicolinerity among the individual library terms.\\
Fig.\ref{fig:cond_VIF_librery} shows the variation of the condition number and the VIF as the data fraction $\varepsilon$ is decreased. The candidate library is already highly ill-conditioned due to strong correlations among its constituent terms. Both the condition number and the VIF, however, increase systematically with decreasing $\varepsilon$, indicating a progressive increase in multicolinerity. As a result, the regression becomes increasingly sensitive to small numerical perturbations, leading to larger deviations in the recovered coefficients, as observed in Fig.~\ref{fig:coeff_diff_data_frac_model}.\\
\subsubsection{Effect of Library Size}
\begin{figure}
    \centering
    \includegraphics[width=0.999\linewidth]{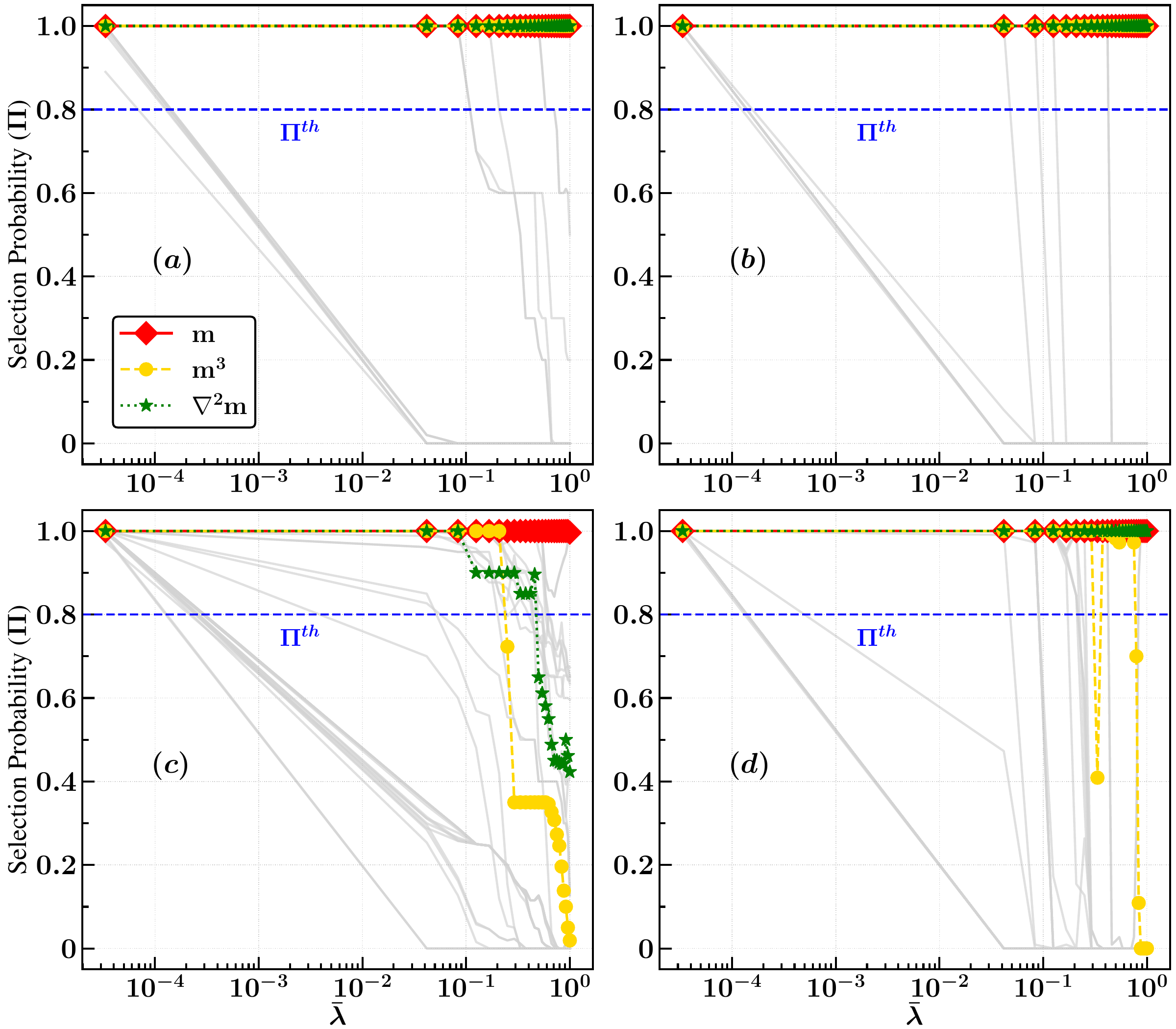}
    \caption{Selection probability, $\Pi$, of the candidate terms is shown as a function of the normalized threshold parameter, $\bar{\lambda}=\lambda/\lambda_{\max}$, for library sizes at different value of $\varepsilon$ : for the $\Theta_{18}$ library at (a) $\varepsilon=0.25$, and (b) $\varepsilon=1.00$; for the $\Theta_{39}$ library at (c) $\varepsilon=0.25$, and (d) $\varepsilon=1.00$. The three terms of the Allen--Cahn equation are highlighted, while the remaining candidate terms are shown in gray. The horizontal dashed line denotes the selection probability threshold, $\Pi^{\mathrm{th}}=0.80$.}
    \label{fig:dif_diff_librery_size}
\end{figure}

Next, we investigate the effect of the candidate-library size on the accuracy of equation recovery by expanding the candidate library to include additional terms. The expanded library contains the spatial derivative terms $(\boldsymbol{\nabla}m)^2$, $\nabla^2m$, $(\boldsymbol{\nabla}m)^2\nabla^2m$, and $\nabla^4m$, together with all allowed products of these terms with polynomial terms in $m$. The resulting terms are restricted to at most fourth order in spatial derivatives and seventh order in $m$. This yields an expanded candidate library containing $39$ terms, denoted by $\Theta_{39}$. We compare the recovered equations obtained using the $\Theta_{18}$ and $\Theta_{39}$ libraries for two representative data fractions, $\varepsilon=0.25$ and $\varepsilon=1.0$. We compare the recovered equations obtained using the $\Theta_{18}$ and $\Theta_{39}$ libraries for two different data fractions, $\varepsilon=0.25$ and $\varepsilon=1.0$, to examine the effect of library size at different levels of data availability.\\
Fig.\ref{fig:dif_diff_librery_size} shows the stability-selection results obtained using the $\Theta_{18}$ and $\Theta_{39}$ libraries for $\varepsilon=0.25$ and $1.0$. Enlarging the candidate library substantially affects the identification of the relevant terms and, consequently, the accuracy of equation recovery. In particular, the selection probabilities of the $m^3$ and $\nabla^2m$ terms, which constitute essential components of the Allen-Cahn equation, decrease markedly for the larger library. In contrast, several additional candidate terms acquire high selection probabilities over a broad range of $\bar{\lambda}$, resulting in the selection of incorrect terms. This degradation can be attributed to the increased multicolinerity introduced by the larger library, which makes it more difficult for the sparse regression to distinguish the true governing terms from strongly correlated alternatives. The effect is further amplified when the amount of available data is reduced, where the combination of a larger library and limited data substantially compromises the efficiency of equation recovery.\\
\begin{figure}
    \centering
    \includegraphics[width=0.99\linewidth]{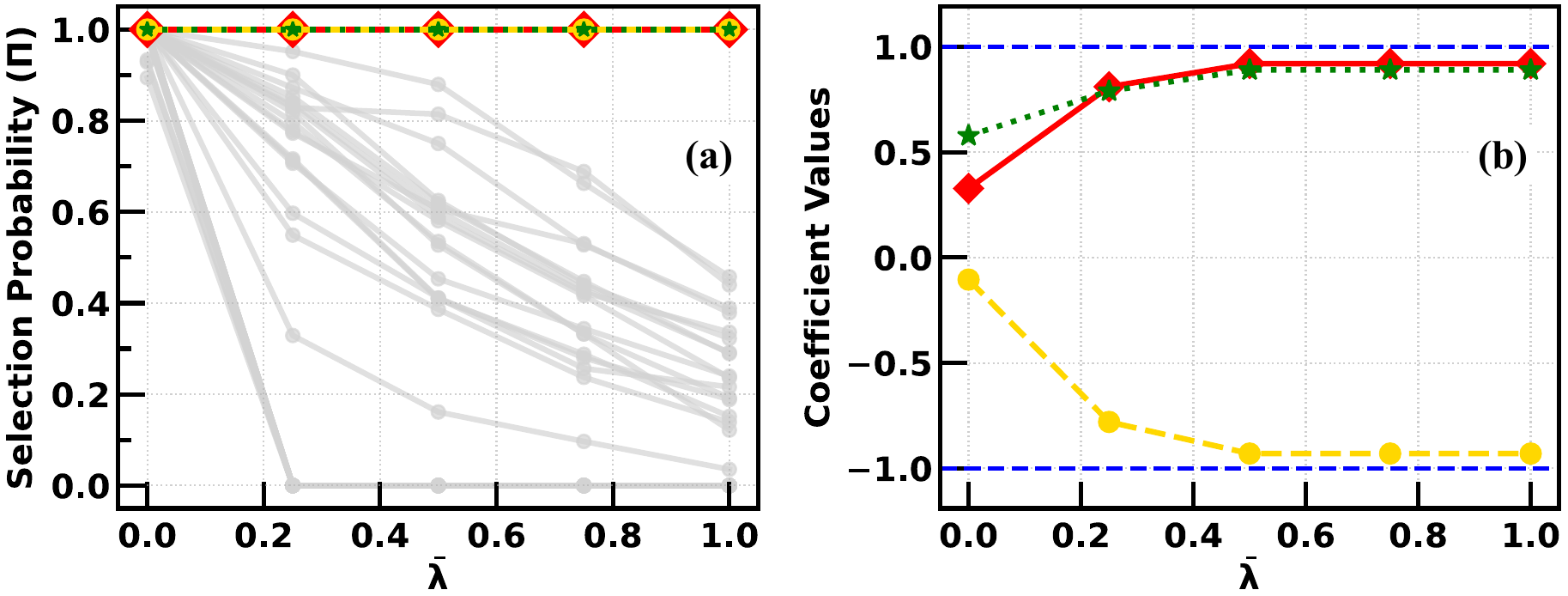}
    \caption{Plot of the selection probability $\Pi$ in (a) for regression with librery bagging for the $\Theta_{39}$ library. Panel (b) shows the coefficient of main three terms. In both panels (a) and (b), the color code is the following: red for $m$, gold for $m^3$, and green for $\nabla^2 m$. In panel (a), rest of the selection probability of the rest of the terms are shown with grey color.}
    \label{fig:all_ch_lib_bag_39}
\end{figure}
Ensemble based variants of SINDy has been proposed to improve the robustness of equation discovery from data by aggregating the information from multiple sub-sampled libraries \cite{fasel2022ensemble}. The strong sensitivity to library size motivates us to examine whether ensemble-based sampling of the candidate library can improve term identification in our case. We therefore incorporate library bagging into the stability-selection PDE-SINDy procedure and apply it to the $\Theta_{39}$ library. The selection probability $\Pi$ for the three terms in Allen-Cahn equation and the corresponding coefficients are shown in Fig.\ref{fig:all_ch_lib_bag_39}(a) and (b), respectively, for $\varepsilon=1.0$. The results with library bagging show an improved separation between the relevant and spurious terms compared with the direct stability-selection results (see Fig.\ref{fig:dif_diff_librery_size}(d)), indicating that library bagging can partially mitigate the ambiguity introduced by the enlarged candidate library. Because of the substantially higher computational cost associated with the larger library, the present analysis (shown in Fig.\ref{fig:all_ch_lib_bag_39}) is based on a smaller number of values of $\bar{\lambda}$ and a limited number of independent regression realizations. The observed improvement should therefore be interpreted as an indication of the potential benefit of library bagging rather than as a comprehensive statistical assessment of its performance.
\subsubsection{Effect of noise}
\begin{figure}
    \centering
    \includegraphics[width=0.69\linewidth]{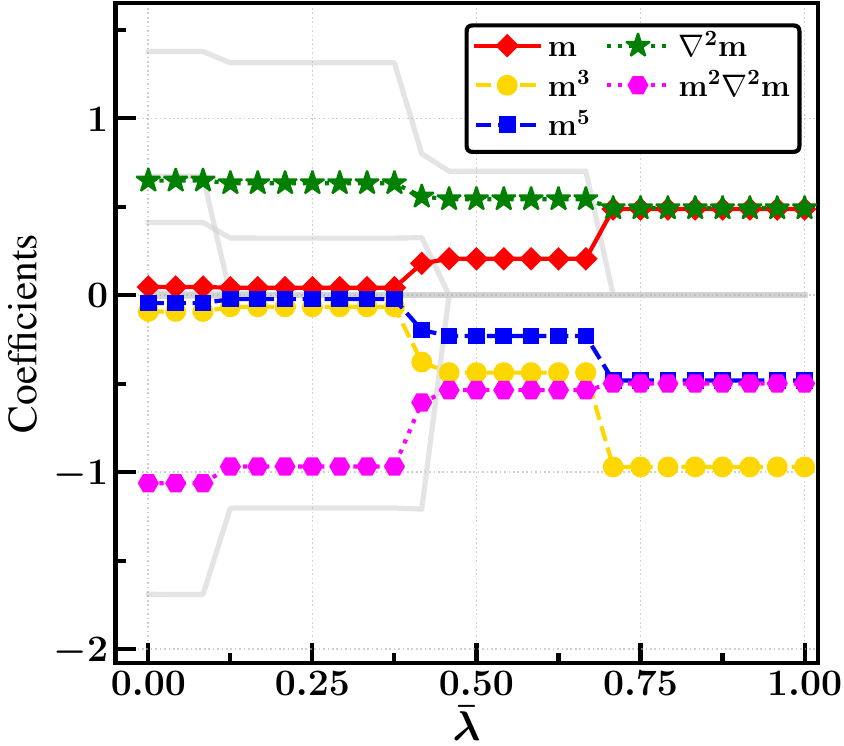}
    \caption{Recovered coefficients of the candidate terms as a function of the normalized threshold parameter $\bar{\lambda}=\lambda/\lambda_{\max}$ for $s=0.01$.}
    \label{fig:coeff_with_noi}
\end{figure}

Finally, we investigate the effect of noise on the accuracy of equation recovery. To generate noisy datasets, additive Gaussian white noise is introduced into the order-parameter field $m(\boldsymbol{r},t)$ according to
\begin{equation*}
\tilde{m}(\boldsymbol{r},t)=m(\boldsymbol{r},t)+\eta(\boldsymbol{r},t)
\end{equation*}
where $\eta(\boldsymbol{r},t)$ denotes a delta-correlated Gaussian random field satisfying
$\langle\eta(\boldsymbol{r},t)\rangle=0$ and $\langle\eta(\boldsymbol{r},t)\eta(\boldsymbol{r}',t')\rangle
=\eta_0\,\delta(\boldsymbol{r}-\boldsymbol{r}')\delta(t-t')$. 
The noise strength is specified as $\eta_0=s\sigma_m$, where $\sigma_m$ denotes the standard deviation of the order-parameter field and $s$ controls the noise amplitude. Such perturbations provide a simple representation of the fluctuations and measurement uncertainties that are inevitably present in experimentally or numerically obtained data. For spin systems, fluctuations arising from the underlying thermal dynamics can contribute to variability in the measured order-parameter field. In experimental measurements, the observed fluctuations may additionally reflect both intrinsic fluctuations of the system and measurement or observation noise.\\
\begin{figure}[hbt]
    \centering
    \includegraphics[width=0.99\linewidth]{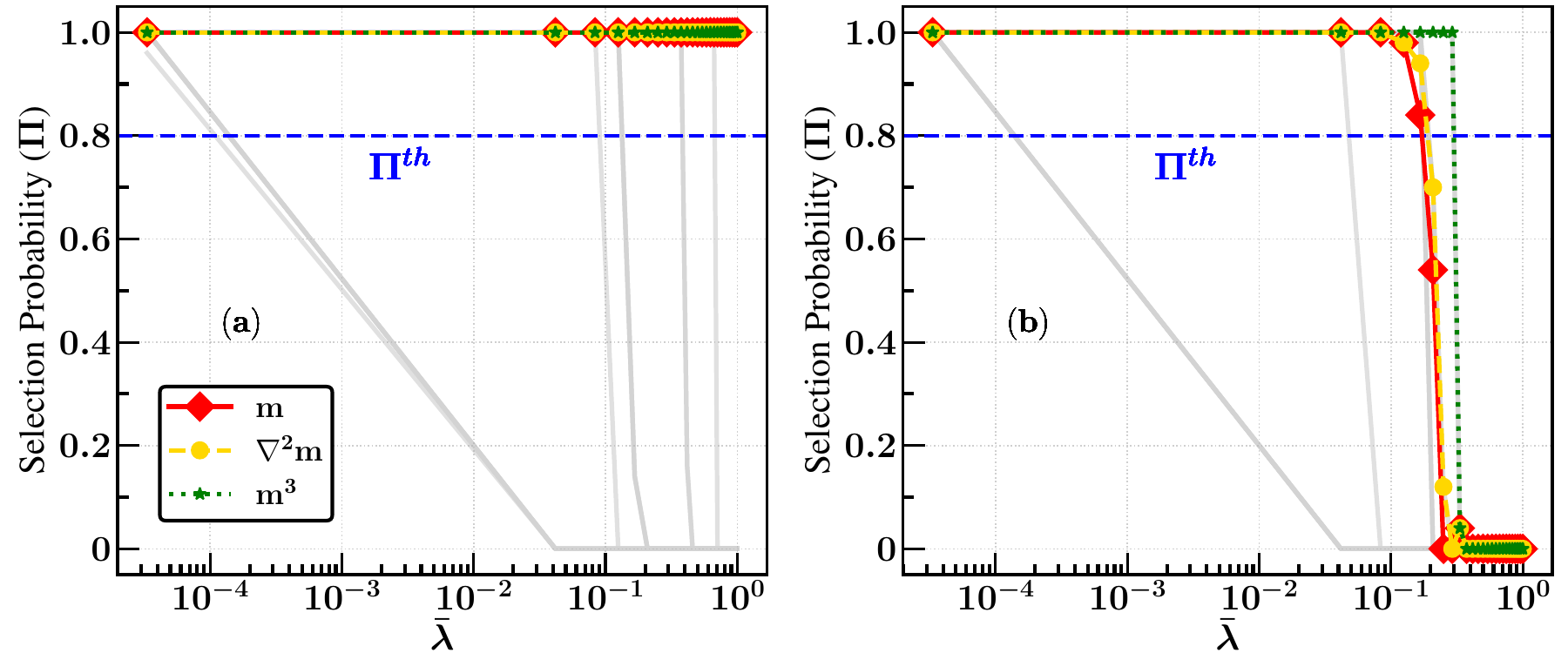}
    \caption{Selection probability ($\Pi$) of the candidate terms as a function of the normalized threshold parameter $\bar{\lambda}=\lambda/\lambda_{\max}$ for the noisy dataset with noise amplitude (a) $s=0.01$ and (b) $s=0.02$. The three true terms in the Allen--Cahn equation, namely $m$, $m^3$, and $\nabla^2m$, are highlighted, while the remaining candidate terms are shown in gray.}
    \label{fig:stability_selection_with_noise}
\end{figure}
We consider two noise levels, $s=0.01$ and $0.02$, and compare the resulting equation-recovery performance with that obtained from the noise-free data. Fig.\ref{fig:stability_selection_with_noise} shows the selection probabilities of the candidate terms for the two noise strengths. For the lower noise level ($s=0.01$), the three terms constituting the Allen--Cahn equation retain unit selection probability over the entire range of $\bar{\lambda}$. However, additional terms, namely $m^5$ and $m^2\nabla^2m$, also acquire high selection probabilities over a finite range of $\bar{\lambda}$. As a result, the inferred model contains spurious terms, preventing the unique identification of the correct governing equation. Upon increasing the noise strength to $s=0.02$, the equation-recovery performance deteriorates further. In particular, for $\bar{\lambda}\geq0.40$, none of the candidate terms satisfies the prescribed selection-probability threshold, indicating that reliable identification of the governing terms becomes increasingly difficult in the presence of stronger noise.\\
Fig.\ref{fig:coeff_with_noi} shows the recovered coefficients for the noise level $s=0.01$. The coefficients of the three terms constituting the Allen--Cahn equation are recovered with the correct signs, but their magnitudes deviate substantially from the corresponding exact values. In addition, the spurious terms $m^5$ and $m^2\nabla^2m$ acquire finite coefficients, further modifying the inferred dynamics from that of the underlying Allen--Cahn equation.\\
These results demonstrate that noise in the data affects both the qualitative identification of the governing equation and the quantitative accuracy of the recovered coefficients, with stronger noise progressively reducing the efficiency of equation recovery.\\
Overall, the results demonstrate that reliable equation recovery requires both robust identification of the relevant terms and accurate estimation of their coefficients, which are influenced by the conditioning of the regression problem and the quality and quantity of the available data. These results provide a basis for assessing equation recovery in more complex dynamical systems.\\
\subsection{Inference of the Coarse-Grained Equation from Glauber Dynamics}
\begin{figure}[hbt]
    \centering
    \includegraphics[width=0.999\linewidth]{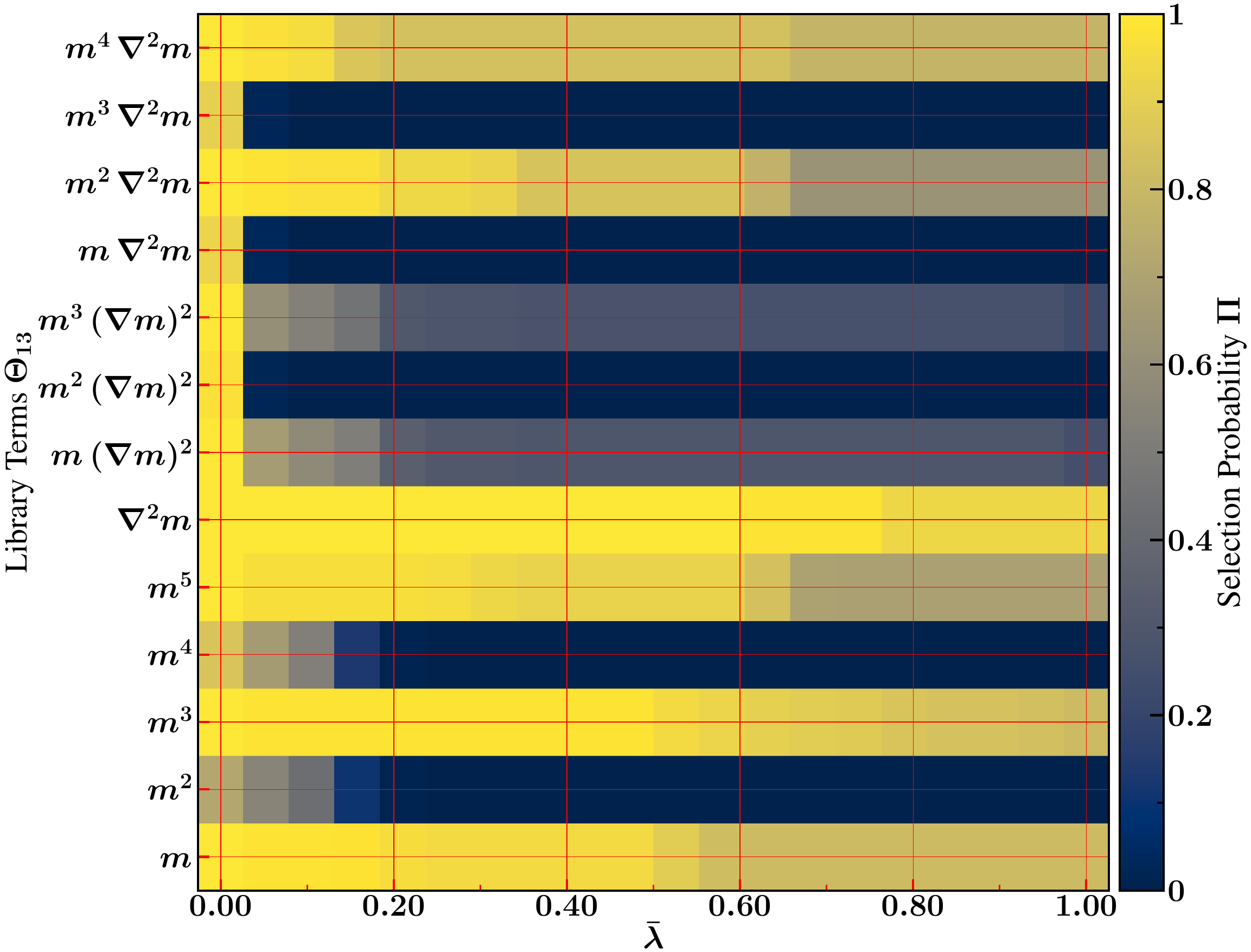}
    \caption{Selection probability, $\Pi$, of the candidate terms in the $18$-term function library, $\Theta_{13}$, obtained from the stability selection PDE-SINDy framework combined with library bagging for the coarse-grained Glauber spin-flip dynamics. The results are shown as a function of the normalized threshold parameter, $\bar{\lambda}=\lambda/\lambda_{\max}$. The color map represents the selection probability of each candidate term.}
    \label{fig:ising_prob_heatmap}
\end{figure}

We next apply the PDE-SINDy framework to infer the coarse-grained dynamics of the Ising model with Glauber spin-flip dynamics (GIM). In contrast to the previous analysis, the governing continuum equation and its coefficients are not known \textit{a priori} and are instead inferred directly from spatiotemporal data generated by Monte Carlo simulation of GIM. The details of the data-generation and coarse-graining procedures are provided in Appendix~\ref{app:gl_ising_model}. For the regression, we employ a $14$-term candidate library, denoted by $\Theta_{13}$, obtained by removing terms of order higher than $\mathcal{O}(m^5,\nabla^2)$ from the $\Theta_{18}$ library. To improve the robustness of the equation-discovery procedure, we combine the stability-selection PDE-SINDy framework with the library-bagging approach, which has been shown to enhance the robustness of sparse regression.\\
Fig.\ref{fig:ising_prob_heatmap} shows the corresponding selection probabilities of the candidate terms. The stability-selection analysis provides a hierarchy of candidate terms according to their selection probabilities, allowing different candidate equations for the coarse-grained magnetization field to be constructed by varying the selection-probability threshold $\Pi^{\mathrm{th}}$. Once the candidate terms are selected, the regression is repeated using only the retained terms to determine their corresponding coefficients. We consider two threshold values, $\Pi^{\mathrm{th}}=0.60$ and $0.80$, representing relatively relaxed and stringent selection criteria, respectively. The resulting equations, denoted as Model-I ($\Pi^{\mathrm{th}}=0.80$) and Model-II ($\Pi^{\mathrm{th}}=0.60$), together with their corresponding coefficients, are shown in Table.\ref{tab:glauber_eq}.\\
\begin{table}[hbt]
\centering
\caption{Recovered coarse-grained equations for the Glauber spin-flip dynamics obtained for two different values of the selection probability threshold, $\Pi^{\mathrm{th}}$.}
\label{tab:glauber_eq}
\renewcommand{\arraystretch}{1.7}
\begin{tabular}{|c|c|c|c|}
\hline
Model & $\Pi^{\mathrm{th}}$ & Recovered Equation & Coefficient Values \\
\hline

I
&
$0.80$
&
\makecell[l]{$\displaystyle
\partial_t m=a_{21}m$\\
$\displaystyle\qquad\;+a_{23}m^3$\\
$\displaystyle\qquad\;
+\kappa_{21}\nabla^2m$}
&
\makecell[l]{
$a_{21}=0.285\pm0.0016$\\
$a_{23}=-0.275\pm0.0016$\\
$\kappa_{21}=0.548\pm0.0025$\\
}
\\
\hline

II
&
$0.60$
&
\makecell[l]{$\displaystyle
\partial_t m=a_{11}m$\\
$\displaystyle\qquad\;
+a_{13}m^3$\\
$\displaystyle\qquad\;
+a_{15}m^5$\\
$\displaystyle\qquad\;
+\kappa_{11}\nabla^2m$\\
$\displaystyle\qquad\;
+\kappa_{12}m^2\nabla^2m$\\
$\displaystyle\qquad\;
+\kappa_{13}m^4\nabla^2m$}
&
\makecell[l]{
$a_{11}=0.0939\pm0.0036$\\
$a_{13}=-0.537\pm0.0012$\\
$a_{15}=-0.620\pm0.0089$\\
$\kappa_{11}=0.606\pm0.0024$\\
$\kappa_{12}=-0.785\pm0.0421$\\
$\kappa_{13}=0.715\pm0.0440$\\
}

\\
\hline
\end{tabular}
\end{table}
\begin{table*}[hbt]
\centering
\renewcommand{\arraystretch}{1.35}
\setlength{\tabcolsep}{14pt}
\caption{Coefficients of the terms in Model-I for different temperatures of the Glauber spin flip Ising model. Values are reported as mean $\pm$ standard deviation.}
\label{tab:model1_coefficients}
\begin{tabular}{|c|c|c|c|}
\hline
$T$ & $a_{21}$ & $a_{23}$ & $\kappa_{21}$ \\
\hline
\hline
0.50 & $0.285 \pm 0.0018$ & $-0.275 \pm 0.0018$ & $0.548 \pm 0.0025$ \\
\hline
0.75 & $0.272 \pm 0.0016$ & $-0.261 \pm 0.0016$ & $0.527 \pm 0.0025$ \\
\hline
1.00 & $0.245 \pm 0.0017$ & $-0.235 \pm 0.0017$ & $0.485 \pm 0.0026$ \\
\hline
1.25 & $0.195 \pm 0.0015$ & $-0.186 \pm 0.0016$ & $0.409 \pm 0.0029$ \\
\hline
\end{tabular}
\end{table*}
\begin{figure}[hbt]
    \centering
    \includegraphics[width=0.999\linewidth]{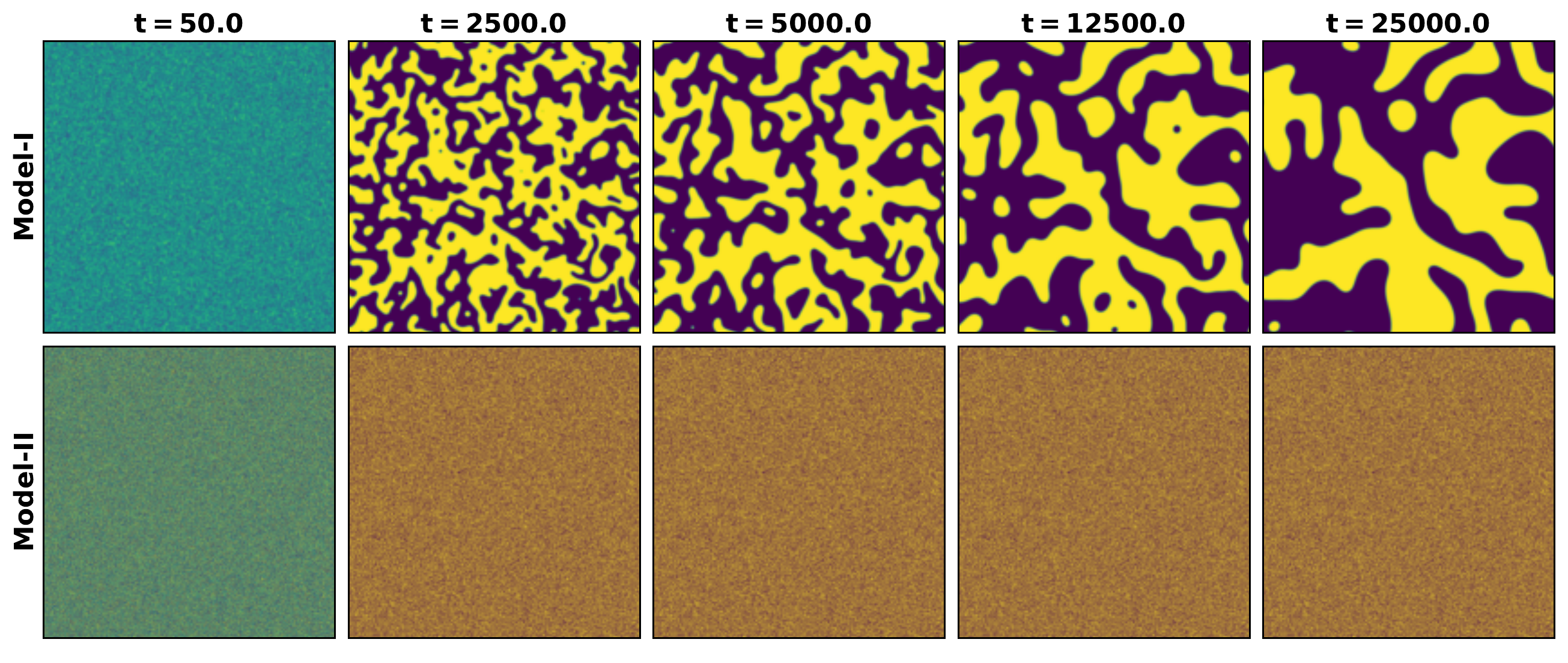}
    \caption{Time evolution of Model-I and II, starting from an homogeneous initial condition. The heatmap shows the local magnetization field $m(\boldsymbol{r},t)$ where the bright (dark) regions represent $m(\boldsymbol{r},t)>0$ ($<0$). The snapshots are shown for lattice size $L=1024$}
    \label{fig:snaps_ising_models_I_II}
\end{figure}
\begin{figure}[hbt]
    \centering
    \includegraphics[width=0.999\linewidth]{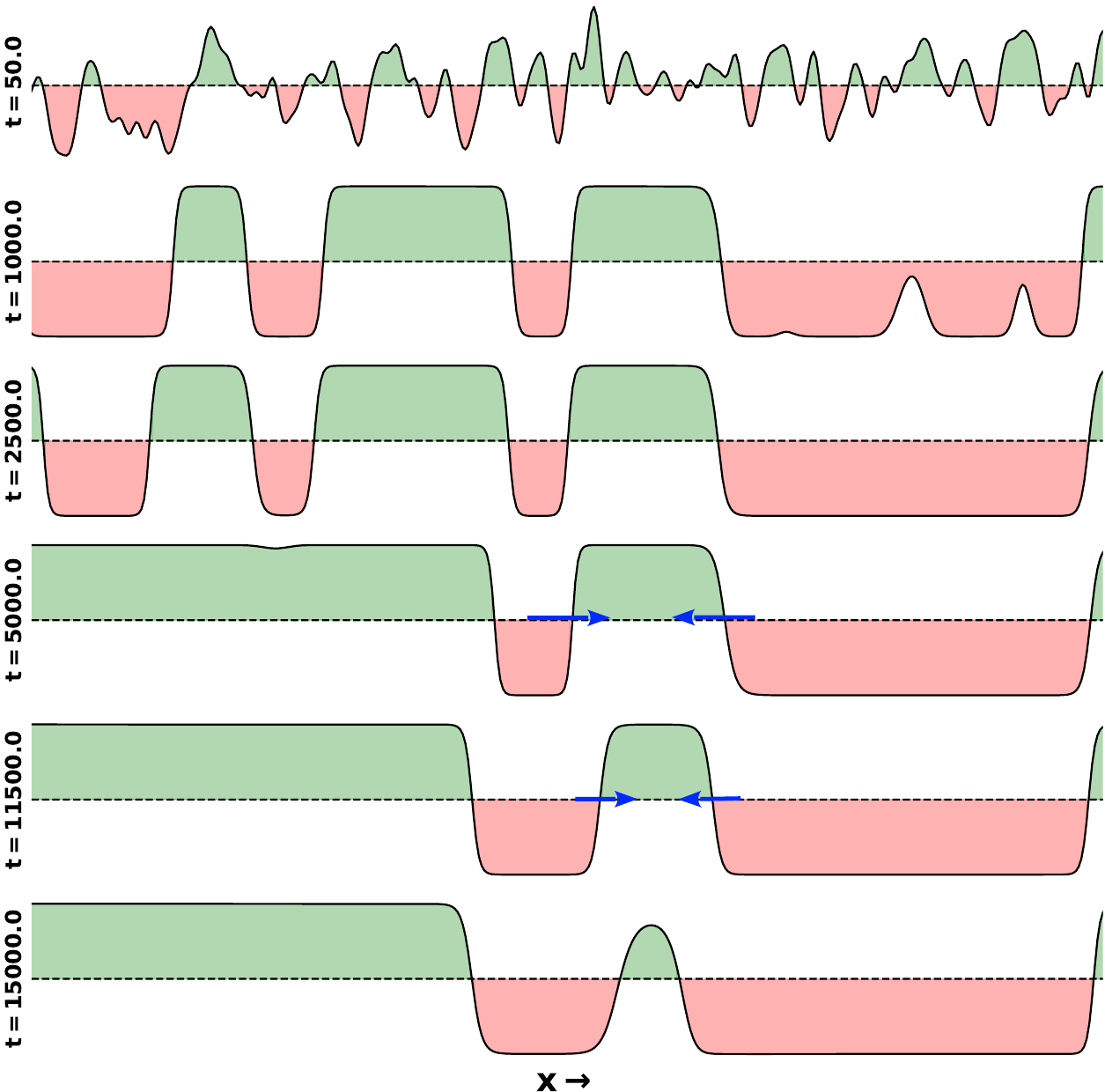}
    \caption{Local curvature driven dynamics of the interfaces and domain coasening in Model-I. The plot shows the 1D plot of magnetization $m(x)$ \emph{vs.} $x$ for $x\in[300,700]$ at $y=512$ for a lattice size of $L=1024$.}
    \label{fig:curvature_driven_dynamics}
\end{figure}
For the more stringent selection threshold, $\Pi^{\mathrm{th}}=0.80$ (Model-I), the recovered equation contains only three terms, $m$, $m^3$, and $\nabla^2m$, and therefore has the same functional form as the leading-order Model-A (Allen-Cahn) equation. This result is consistent with the established correspondence between the coarse-grained dynamics of the Ising model with Glauber spin-flip dynamics and the Model-A universality class. Moreover, the signs and relative magnitudes of the recovered coefficients are consistent with the mean-field analysis of the inferred equation, which predicts phase separation with stable bulk magnetization values $m_0=\pm\sqrt{-a_{21}/a_{23}}$.\\
The coefficients identified by PDE-SINDy for Model-I at different temperatures are summarized in Table~\ref{tab:model1_coefficients}. The same three terms, $m$, $m^3$, and $\nabla^2m$, are consistently recovered with unchanged signs across the investigated temperature range. The magnitudes of all three coefficients decrease systematically with increasing temperature, with the coefficients of the $m$ and $m^3$ terms exhibiting a similar temperature dependence. The small standard deviations indicate good reproducibility of the recovered coefficients across the regression realizations. Thus, while increasing temperature modifies the effective coefficients of the coarse-grained equation, its functional form remains unchanged.\\
To validate the emergent dynamics described by the inferred equations, we simulate the recovered equations corresponding to Model-I and Model-II using a lattice spacing of $\Delta x=1.0$ and a time step of $\Delta t=0.01$, and compare their resulting dynamics with those of the original Ising system. For both models, the simulations are initialized from a random homogeneous initial condition. Fig.\ref{fig:snaps_ising_models_I_II} presents the temporal evolution of the two recovered models on a $1024\times1024$ lattice.\\
The snapshots reveal a pronounced qualitative difference between the dynamics predicted by the two models. Model-II does not exhibit phase separation, indicating that the additional terms retained under the relaxed selection criterion qualitatively alter the dynamics of the system and prevent the emergence of the coarsening state characteristic of the Ising system below the critical temperature. In contrast, Model-I develops well-defined domains that grow with time through curvature-driven interface motion (see Fig.\ref{fig:curvature_driven_dynamics}), in qualitative agreement with the dynamics of the original Ising model.\\
To further quantify the agreement between the coarsening dynamics of Model-I and that in GIM, we calculate the temporal evolution of the characteristic domain length scale, $l(t)$. The details of the calculation of $l(t)$ are provided in Appendix~\ref{app:gr_kin_GIM}. As shown in Fig.~\ref{fig:len_Model-I}, the characteristic length scale follows the scaling $l(t)\sim t^{1/2}$, consistent with the expected curvature-driven coarsening kinetics of the Glauber Ising model. This agreement further demonstrates that Model-I faithfully captures the coarsening dynamics of the underlying microscopic system.

\section{Discussion}
In this work, we have investigated the applicability of the PDE-SINDy framework to phase-ordering systems using spatiotemporal data. We first considered the Allen--Cahn equation as a controlled benchmark and subsequently extended the analysis to the Ising model with Glauber spin-flip dynamics. This provides a systematic setting to assess the challenges associated with data-driven discovery of coarse-grained dynamical equations for phase-ordering systems.\\ 
An important outcome of our study is the identification of several factors that influence the efficiency of data-driven equation discovery. (i) First, our results reveal a clear distinction between qualitative and quantitative recovery: the stability-selection PDE-SINDy framework can robustly identify the relevant terms in the governing equation even when the available data are limited or noisy, whereas the corresponding coefficient values can exhibit substantially larger deviations. (ii) Second, the accuracy of both term identification and coefficient recovery is sensitive to the size of the candidate function library. This highlights the importance of carefully constructing the library using physical constraints, such as symmetry and conservation laws, and, where appropriate, restricting it to the leading-order terms expected from the underlying dynamics. (iii) Third, our results demonstrate that ensemble-based approaches, such as library bagging, can improve the robustness of equation discovery by reducing the sensitivity to a particular choice of candidate library. Finally, combining library bagging with stability-selection PDE-SINDy enables the identification of a coarse-grained dynamical equation for the Ising model with Glauber spin-flip dynamics that reproduces the characteristic phase-separation and coarsening dynamics of the underlying microscopic system.\\
The present analysis also points to several directions in which data-driven equation discovery can be extended to more complex dynamical systems. Natural extensions include conserved order-parameter dynamics \cite{hohenberg1977theory}, coupled and multicomponent fields \cite{chen2002phase}, and vector or tensorial order parameters \cite{de1995physics}, as well as nonequilibrium and active systems \cite{RevModPhys.85.1143}, where the relevant continuum descriptions may involve complex nonlinearities and couplings. Even though data-driven equation discovery has been applied to several of these systems \cite{PhysRevLett.129.258001,golden2023physically,supekar2023learning,schaeffer2017learning}, a systematic analysis of the factors limiting term identification and coefficient recovery remains an important direction for future work. Further, the robustness of equation discovery could be improved by incorporating approaches such as weak-form formulations of SINDy \cite{messenger2021weak}, or other variants of SINDy \cite{kaiser2018sparse,fasel2022ensemble}, which are less sensitive to measurement noise and the numerical errors associated with direct differentiation of noisy data. For such systems, where multiple symmetry-allowed terms may be strongly correlated, physical constraints and dynamical validation will become increasingly important for distinguishing between competing candidate equations. In particular, comparing the predicted stability, coarsening behavior, or other relevant dynamical observables with those of the underlying system can provide an additional criterion for selecting among statistically plausible models.

\section*{Acknowledgement}
P.S.M. thanks UGC for the research fellowship. A.K. thanks PMRF, INDIA, for the research fellowship. S.M. thanks DST, SERB (INDIA), Project No.: CRG/2021/006945, MTR/2021/000438, and ANRF grant numbered ANRF/ARG/2025/008220/PS  for financial support.\\

\textbf{Author contributions:} P.S.M., M.K.J., A.K., and S.M. conceptualized the project; M.K.J. performed the initial checks; P.S.M. designed the SINDY algorithm, carried out the simulations, and performed the initial data analysis; P.S.M., M.K.J., A.K., and S.M. interpreted and analyzed the results; P.S.M. prepared the first draft of the manuscript; All authors reviewed, edited, and approved the final version of the manuscript. S.M. supervised the project.

\onecolumngrid
\appendix
\section{Methods and Algorithm Details}\label{app:meth_alg_det}
\subsection{Data Generation}\label{app:data_gen_ac_gim}
In this section we provide details of simulation and data generation from the Allen-Cahn equation and the Glauber Ising model in 2-dimensions.

\subsubsection{Allen-Cahn Equation}\label{app:allen_cahn}
The Allen-Cahn equation for scalar order parameter field $m(\boldsymbol{r},t)$ is given by \cite{hohenberg1977theory,bray1994theory}
\begin{equation}
    \frac{\partial m}{\partial t} = \alpha m + \beta m^3 + \kappa\nabla^2m
    \label{eq:allencahn}
\end{equation}
where, $\beta<0$ and $\kappa>0$. Depending on the sign of $\alpha$ we get a homogeneous state ($\alpha<0$) or a phase separated state ($\alpha>0$).\\
Eq.\ref{eq:allencahn} is simulated on a $2$-dimensional square lattice of size $256\times256$  with periodic boundary condition in both $X$ and $Y$-directions with $\Delta x=1.0$ and $\Delta t=0.01$. The spatial derivatives are calculated using the $9$-point central difference scheme and the temporal evolution is performed using the forward Euler scheme. The system is simulated for $10^5$ simulation steps. One simulation step is counted when $m(\boldsymbol{r},t)$ is updated once at every lattice point.\\
In our simulations, the parameters in Eq.\ref{eq:allencahn} are fixed at $(\alpha,\beta,\kappa) = (1,-1,1)$ and for an initial configuration with $\left\langle m \right\rangle_{\boldsymbol{r}} = 0$. To create the dataset, we have considered $10$-distinct trajectories of the system starting from independent random initial conditions, and for each trajectory $500$ snapshots are stored after first $500$ steps at the gap of $200$ time steps.

\subsubsection{Glauber Ising Model}\label{app:gl_ising_model}
We consider the $2$-dimensional Ising model defined on a square lattice of size $1024 \times 1024$ with periodic boundary conditions along both $X$ and $Y$-direction. The system consists of spins ${s_i}$ located at each lattice site $i$. Each spin variable can take one of the two possible values, $s_i = \pm 1$, corresponding respectively to the up and down orientations of the spin. The interaction between spins favors local alignment, such that neighboring spins tend to orient parallel to each other in order to minimize the energy of the system. The corresponding Hamiltonian is given by, \cite{yeomans1992statistical,huang2008statistical}
\begin{equation}
H = -J \sum_{\langle i,j \rangle} s_i s_j
\end{equation}
where $J > 0$ denotes the ferromagnetic coupling and the summation $\langle i,j \rangle$ runs over all nearest-neighbor pairs on the lattice. For positive values of $J$, configurations with parallel neighboring spins are energetically favored, leading to the emergence of long-range ferromagnetic order below the critical temperature in two dimensions.\\
We begin with a completely disordered initial configuration in which the spin at each lattice site independently takes the value $s_i=\pm1$ with equal probability, corresponding to the high-temperature phase of the system above the critical temperature $T_c$. The system is subsequently quenched to a temperature $T<T_c$ and allowed to evolve dynamically toward an ordered state. The temporal evolution of the spin configurations is simulated numerically using Markov Chain Monte Carlo (MCMC) dynamics \cite{landau2021guide}. In each update step, a trial move is proposed and the corresponding change in energy is calculated as $\Delta E = H_{\mathrm{final}} - H_{\mathrm{initial}}$. The trial move is accepted with probability $P = \min\left(1,e^{-\beta \Delta E}\right)$.where $\beta = \frac{1}{k_B T}$ denotes the inverse temperature and $k_B$ is the Boltzmann constant. One Monte Carlo step (MCS) corresponds to $L^2$ attempted updates of the system.\\
To create the data set, we generate $10$-distinct trajectories of the system and for each trajectory $800$ snapshots are stored after first $50$ MCS at the gap of $4$ MCS. From the snapshots of the spins configurations, configurations of the local magnetization field $m(\boldsymbol{r},t)$ are obtained using the block coarse-graining procedure such that
\begin{equation}
m(\boldsymbol{r},t)=\frac{1}{N_b}\sum_{i \in \mathcal{B}(\boldsymbol{r})} s_i(t)
\end{equation}
where $\mathcal{B}(\boldsymbol{r})$ denotes the coarse-graining block centered around position $\boldsymbol{r}$ and $N_b$ represents the total number of spins contained within the block.\\

\subsection{PDE-SINDy framework}\label{app:pde_sindy}
The governing evolution equation for a field $m(\mathbf{r},t)$ can, in general, be written as
\begin{equation}
\frac{\partial m}{\partial t}=f\left(m,\nabla m,\nabla^{2}m,\ldots\right)
\end{equation}
where $f$ is an unknown function of the field and its spatial derivatives that governs the dynamics of the system. In the PDE-SINDy framework, $f$ is assumed to admit a sparse representation in terms of a predefined library of candidate functions,
\begin{equation}
f=\Theta\boldsymbol{\xi}
\end{equation}
where $\Theta$ is the library constructed from the field $m(\mathbf{r},t)$ and its spatial derivatives, and $\boldsymbol{\xi}$ is the corresponding coefficient vector. The function library $\Theta$ is constructed by first identifying a set of possible functions of the field $m(\boldsymbol{r},t)$ based on the understanding of the system and the domain knowledge. On top of that, the set of possible function can be restricted using the idea of the symmetry and conservation laws of the system.\\
\textbf{\uline{\emph{Regression setup}} :} The field $m(\boldsymbol{r},t)$ is defined on an $L\times L$ lattice over $N_t$ time snapshots. The temporal derivative field $m_t(\mathbf{r},t)$ is computed using the forward difference scheme at all spatial points and time instances, and subsequently flattened into the column vector
\begin{equation}
\mathbf{M}_t^{(m)}
=
\begin{bmatrix}
m_t(\mathbf{r}_1,t_1) \\
m_t(\mathbf{r}_2,t_1) \\
\vdots \\
m_t(\mathbf{r}_{L^2},t_{N_t})
\end{bmatrix}
\end{equation}
containing a total of $L^2N_t$ entries.\\
Similarly, each candidate function in the library is evaluated using a 9-point central difference scheme over all spatial points and time snapshots, flattened into a column vector, and assembled as a column of the library matrix,
\begin{equation}
\Theta^{(m)}
=
\begin{bmatrix}
| & | & & |\\
\Theta_1^{(m)} & \Theta_2^{(m)} & \cdots & \Theta_N^{(m)}\\
| & | & & |
\end{bmatrix}
\end{equation}
where $N$ denotes the total number of candidate functions in the library.\\
The discovery of the governing equation is thereby transformed into a sparse regression problem of determining the coefficient vector $\boldsymbol{\xi}$ such that
\begin{equation}
\mathbf{M}_t=\Theta\boldsymbol{\xi}
\end{equation}
where $\boldsymbol{\xi}$ is obtained using sparse regression techniques.\\
Prior to regression, each column of the library matrix $\Theta$ is normalized independently to eliminate numerical bias arising from the different magnitudes of the candidate functions.\\
The sparse regression problem is then solved using the Sequential Threshold Ridge Regression (STRidge) algorithm. STRidge iteratively eliminates the insignificant terms by performing ridge regression followed by thresholding of the coefficients below a prescribed cutoff value $\lambda$, until convergence. Further to improve the robustness of the discovered equation, we employ stability selection by performing the ensemble regression over multiple subsampled datasets. Additionally for the Glauber Ising model, library bagging is incorporated to improve the efficiency of equation recovery in the presence of noisy data. The details of stability selection and library bagging are provided in appendix B.

\subsection{Algorithm Details}\label{app:alg_det}
\subsubsection{Sampling of snapshots to create different data fractions}\label{app:alg_det_data_sam}
To investigate the effect of data availability, datasets corresponding to different data fractions, $\varepsilon$, are generated by randomly selecting $\varepsilon N_{\mathrm{tot}}$ consecutive snapshots from the complete trajectory containing $N_{\mathrm{tot}}$ snapshots. The selected snapshots are used to evaluate the spatial and temporal derivatives and to construct the corresponding candidate library matrix. Since a single realization at small values of $\varepsilon$ samples only a limited portion of the complete trajectory, the sampling procedure is repeated multiple times with different randomly chosen starting locations. Consequently, the regression is performed over several independently sampled subsets, ensuring that the complete trajectory contributes to the statistical analysis. The number of sampling realizations is increased with decreasing $\varepsilon$ to maintain adequate coverage of the available data.

\subsubsection{Stability selection for the Allen--Cahn equation}\label{app:alg_det_stab_sel}
After constructing the regression problem, $\partial_t m=\Theta\xi$, stability selection is performed through repeated subsampling of the available observations. For each realization, $50\%$ of the rows of the regression matrix $\Theta$ are selected at random without replacement, and sparse regression is performed on the resulting subproblem ($\partial_t m^{\mathrm{sub}}=\Theta^{\mathrm{sub}}\xi^{\mathrm{sub}}$) for a prescribed value of the threshold parameter $\lambda$. This procedure is repeated $1000$ times for each value of $\lambda$. The selection probability of the $i^{\mathrm{th}}$ candidate function is then defined as $\Pi_i=\frac{N_i^{\mathrm{sel}}}{1000}$, where $N_i^{\mathrm{sel}}$ denotes the number of realizations in which the corresponding term is retained after the regression. The resulting selection probabilities provide a measure of the statistical robustness of each candidate term against variations in the sampled data.

\subsubsection{Stability selection with library bagging for the Glauber spin-flip Ising model}\label{app:alg_det_stab_sel_lib_bag}
For the Glauber spin-flip Ising model, stability selection is combined with library bagging to further improve the robustness of equation discovery. In each library-bagging realization, $70\%$ of the candidate functions are randomly selected from the complete library to construct a reduced regression problem. Stability selection is then performed on the reduced library following the procedure described above. For the $\Theta_{14}$ candidate library, $150$ independent library-bagging realizations are generated, and for each sampled library, stability selection is carried out using $500$ independent data-subsampling realizations. The final selection probability of each candidate function is obtained by aggregating the stability-selection results over all library-bagging realizations, thereby reducing the influence of highly correlated candidate functions and improving the robustness of the inferred continuum equation.

\section{Stability selection results for different data fractions}\label{app:stab_sel_res_diff_data_frac}
\begin{figure}[hbt!]
    \centering
    \includegraphics[width=0.999\linewidth]{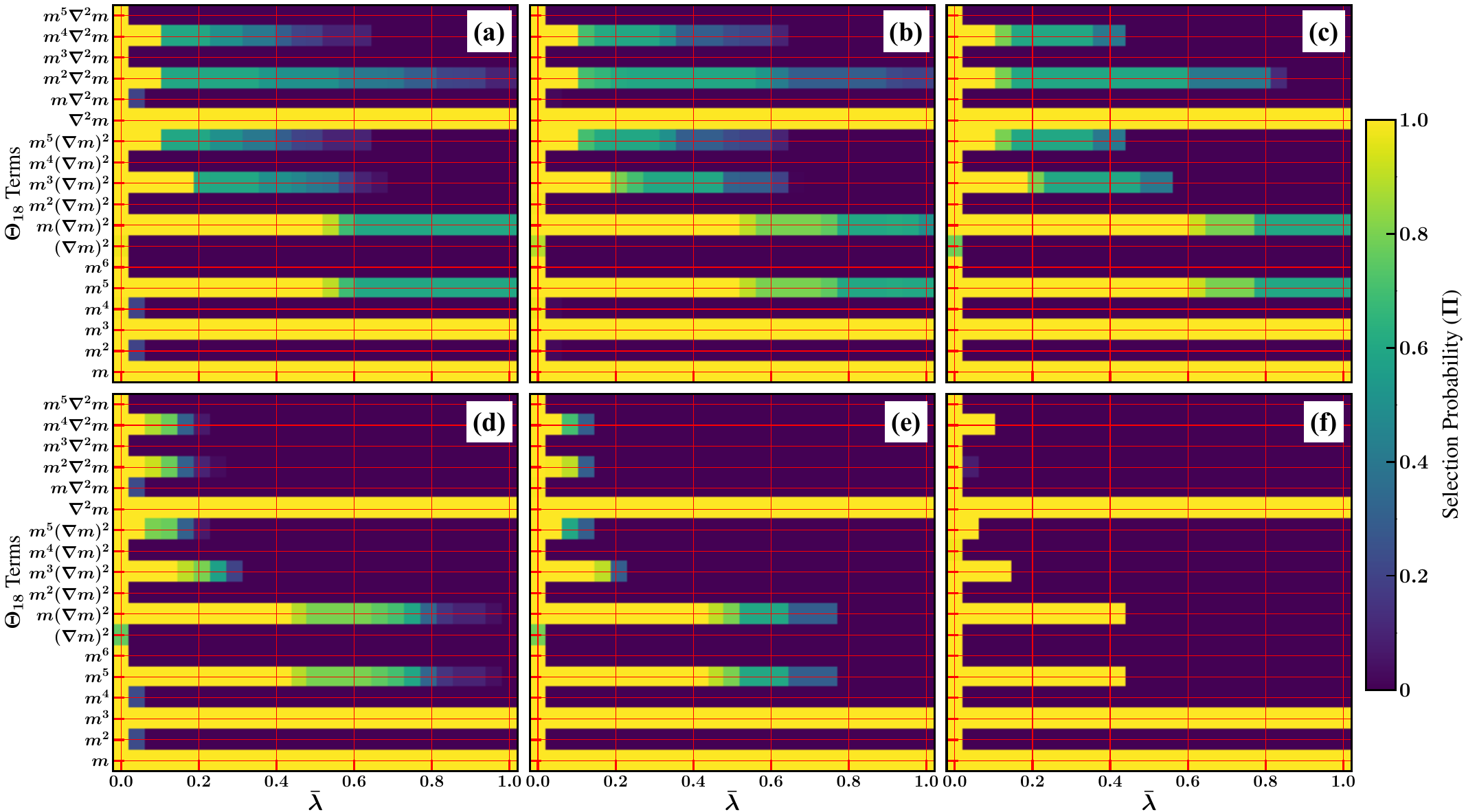}
    \caption{Selection probability ($\Pi$) of all candidate terms in the $18$-term library, $\Theta_{18}$, as a function of the normalized threshold parameter $\bar{\lambda}=\lambda/\lambda_{\max}$ for different fractions of the available data: (a) $\varepsilon=0.10$, (b) $\varepsilon=0.25$, (c) $\varepsilon=0.50$, (d) $\varepsilon=0.75$, (e) $\varepsilon=0.90$, and (f) $\varepsilon=1.00$. The color scale represents the selection probability.}
    \label{fig:heatmap_diff_data_frac}
\end{figure}

Fig.\ref{fig:heatmap_diff_data_frac} presents the complete stability-selection maps for the $18$-term candidate library corresponding to different data fractions. Each row represents a candidate term in the library, while the color indicates its selection probability as a function of the normalized threshold parameter, $\bar{\lambda}$. The three terms constituting the Allen--Cahn equation, namely $m$, $m^3$, and $\nabla^2m$, retain unit selection probability over a wide range of $\bar{\lambda}$ for all values of $\varepsilon$. In contrast, several spurious terms also attain high selection probabilities, particularly for smaller data fractions. As the amount of available data increases, the selection probabilities of these spurious terms decrease progressively, resulting in a clearer separation between the relevant and irrelevant candidate functions. These results are consistent with the observations discussed in Sec.~III A.\\
\section{Stability selection results for different library size}\label{app:stab_sel_res_diff_lib_size}
\begin{figure}[hbt!]
    \centering
    \includegraphics[width=0.995\linewidth]{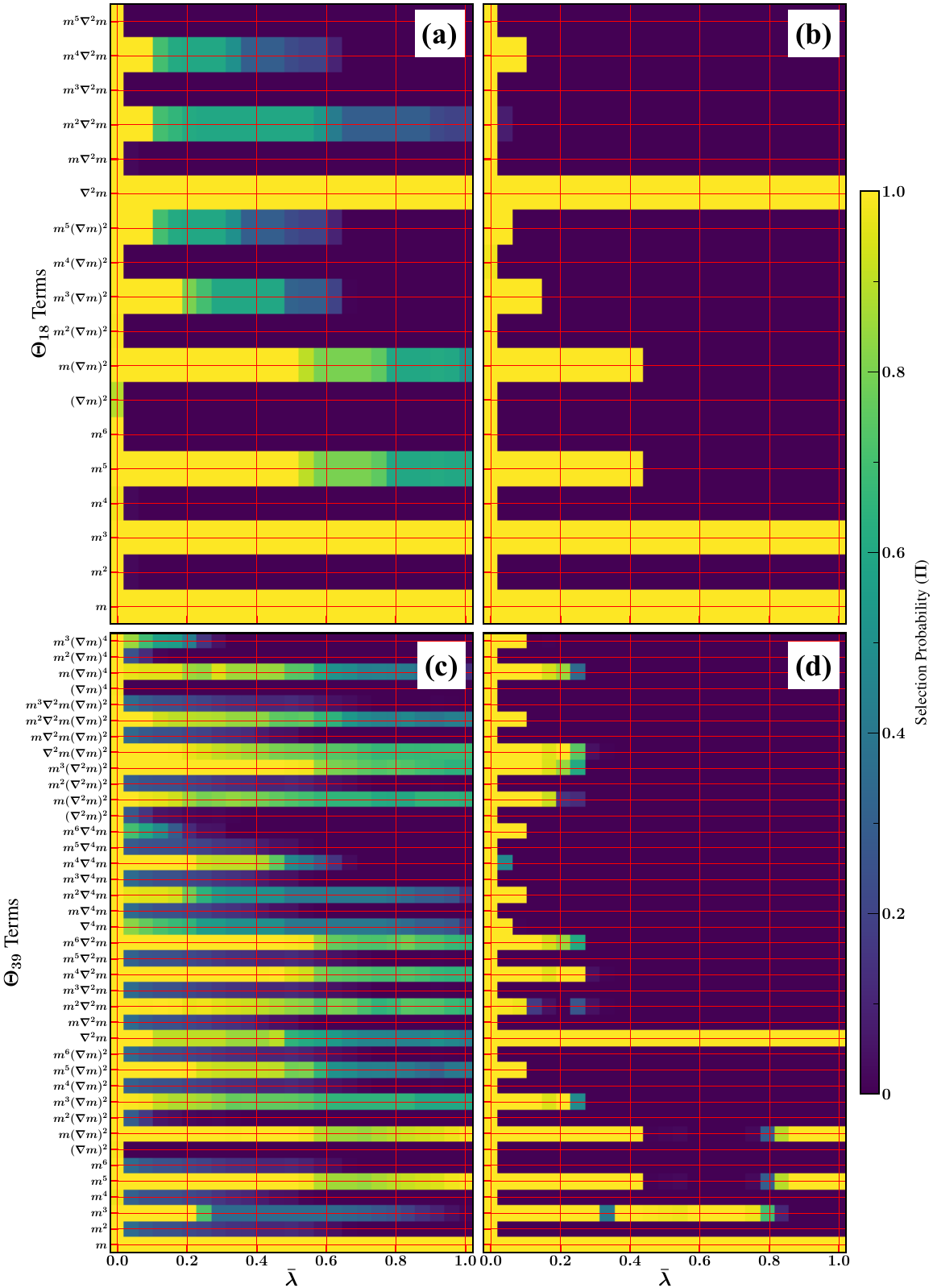}
    \caption{Complete stability-selection maps for the $20$- and $39$-term candidate libraries. Panels (a) and (b) correspond to the $20$-term library, $\Theta_{18}$, while panels (c) and (d) correspond to the $39$-term library, $\Theta_{39}$. The results are shown for two different data fractions: $\varepsilon=0.25$ [(a),(c)] and $\varepsilon=1.00$ [(b),(d)]. The color scale denotes the selection probability, $\Pi$, as a function of the normalized threshold parameter $\bar{\lambda}=\lambda/\lambda_{\max}$. Compared to $\Theta_{18}$, the larger candidate library exhibits a significantly larger number of competing terms with high selection probability, particularly for $\varepsilon=0.25$, illustrating the increased difficulty of correctly identifying the governing equation in the presence of an enlarged function library and limited data.}
    \label{fig:stability_heatmap_diff_lib_size}
\end{figure}

Fig.\ref{fig:stability_heatmap_diff_lib_size} presents the complete stability-selection maps for the $18$- and $39$-term candidate libraries at two different data fractions. The results clearly show that enlarging the candidate library substantially increases the number of competing terms with high selection probability. Consequently, the distinction between the relevant and irrelevant terms becomes less pronounced, increasing the probability of incorrect model selection. This effect is particularly severe for the larger library at $\varepsilon=0.25$, where several additional terms attain selection probabilities comparable to those of the true terms over a broad range of $\bar{\lambda}$. In contrast, for $\varepsilon=1.00$, most of these competing terms are effectively suppressed, resulting in a clearer separation between the relevant and spurious candidate functions. These observations corroborate the discussion in Sec.~III~B regarding the combined influence of library size and data availability on the robustness of equation recovery.
\section{Stability selection results for noisy data}\label{app:stab_sel_res_noise}
\begin{figure}[hbt!]
    \centering
    \includegraphics[width=0.999\linewidth]{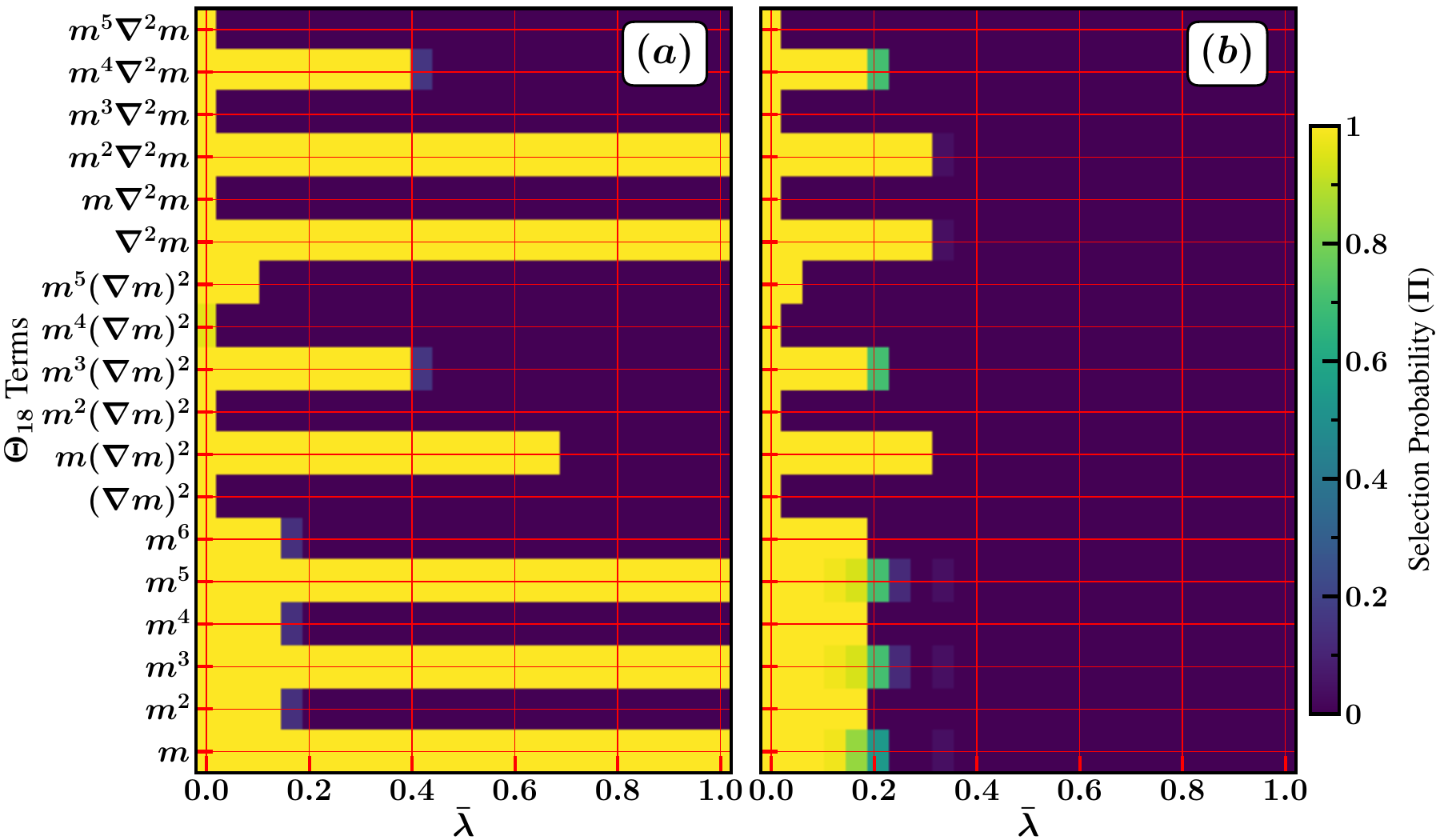}
    \caption{Stability selection heatmaps showing the selection probability, $\Pi$, of the candidate functions in the $\Theta_{18}$ library as a function of the normalized threshold parameter, $\bar{\lambda}=\lambda/\lambda_{\max}$, for two different noise strengths: (a) $s=0.01$ and (b) $s=0.02$. The color represents the selection probability of each library term.}
    \label{fig:noise_stability_heatmap}
\end{figure}

Fig.\ref{fig:noise_stability_heatmap} presents the complete stability selection heatmaps for the $\Theta_{18}$ library at two different noise strengths. For the lower noise level ($s=0.01$), the three governing terms, $m$, $m^3$, and $\nabla^2m$, remain stable over a relatively broad range of the threshold parameter. However, a few higher-order terms, such as $m^5$ and $m(\nabla m)^2$, also attain appreciable selection probabilities over a limited range of $\bar{\lambda}$, indicating that noise promotes the appearance of weakly stable spurious terms. As the noise strength is increased to $s=0.02$, the selection probabilities of all candidate terms decrease substantially, and the range of $\bar{\lambda}$ over which the true governing terms remain stable is significantly reduced. These results illustrate that increasing noise not only weakens the selection probability of the correct terms but also diminishes the overall robustness of the equation discovery process.

\section{Growth Kinetics results from the Derived equation for GIM}\label{app:gr_kin_GIM}
\begin{figure}[hbt]
    \centering
    \includegraphics[width=0.95\linewidth]{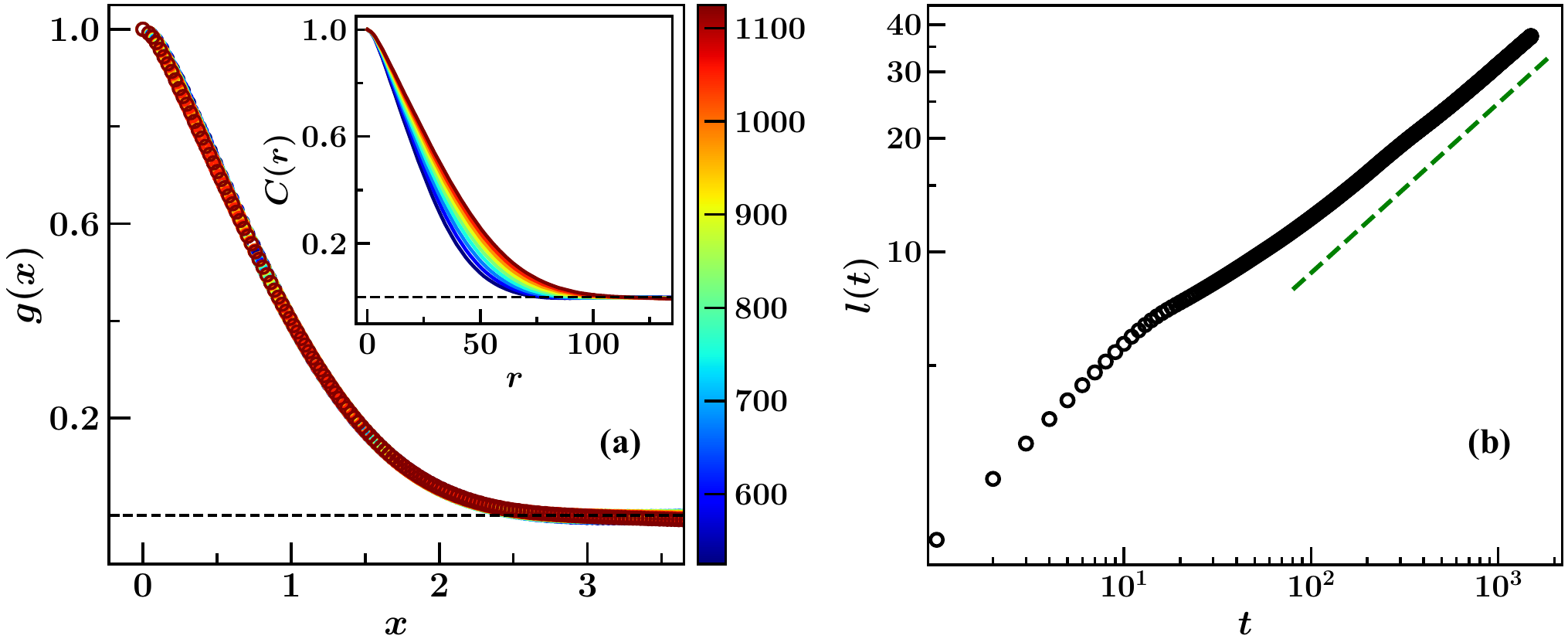}
    \caption{(a) The plot of the scaled correlation function $g(x)$ \emph{vs.} $x=r/l(t)$ at different times. (Inset) The corresponding unscaled correlation functions $C(r)$ \emph{vs.} $r$.(b)The plot of the time evolution of the characteristic length scale $l(t)$ plot for Model-I. The green dashed line marks the slope $t^{1/2}$.}
    \label{fig:len_Model-I}
\end{figure}

To characterize the domain-growth kinetics of Model-I, recovered from the Glauber spin-flip dynamics of the Ising model, we calculate the equal-time two-point correlation function of the magnetization field $m(\boldsymbol{r},t)$,
\begin{equation}
    C(r,t)=\left\langle m(\boldsymbol{r}_0,t)
    m(\boldsymbol{r}_0+\boldsymbol{r},t)\right\rangle,
\end{equation}
where $\langle\cdots\rangle$ denotes an average over the reference position $\boldsymbol{r}_0$, the angular orientation of $\boldsymbol{r}$, and independent realizations. The correlation function is normalized by its value at $r=0$. The plot of the correlation function $C(r)$ at different times is shown in the inset of Fig.\ref{fig:len_Model-I}(a). We define the characteristic domain length scale $l(t)$ as the distance at which the normalized correlation function first decays to $0.5$. At late times, the characteristic length scale grows as a power law as $l(t) \sim t^{1/z}$ with growth exponent $1/z \approx 0.50$ as shown in Fig.\ref{fig:len_Model-I}(b). Further, the correlation functions at different times collapse on a single master curve when distance is scaled as $r to x=r/l(t)$ as shown in the main panel of Fig.\ref{fig:len_Model-I}(a).





\newpage
\twocolumngrid
\bibliography{references}

\end{document}